\documentclass{article}
\usepackage[nonatbib,final]{neurips_2022}

\usepackage{amsmath,amsfonts,bm}

\def\eqref#1{equation~\ref{#1}}

\def\1{\bm{1}}

\DeclareMathAlphabet{\mathsfit}{\encodingdefault}{\sfdefault}{m}{sl}
\SetMathAlphabet{\mathsfit}{bold}{\encodingdefault}{\sfdefault}{bx}{n}

\usepackage{color,xcolor}
\usepackage{epsfig}
\usepackage{graphicx}

\usepackage{adjustbox}
\usepackage{array}
\usepackage{booktabs}
\usepackage{colortbl}
\usepackage{wrapfig}
\usepackage{hhline}
\usepackage{multirow}
\usepackage{subcaption} %
\usepackage{wrapfig}
\usepackage{floatflt}

\usepackage{amsmath,amsfonts,amssymb}
\usepackage{mathtools}  %
\usepackage{bm}
\usepackage{nicefrac}
\usepackage{microtype}

\usepackage{changepage}
\usepackage{extramarks}
\usepackage{fancyhdr}
\usepackage{lastpage}
\usepackage{setspace}
\usepackage{soul}
\usepackage{xspace}

\usepackage[pagebackref=true,breaklinks=true,colorlinks,citecolor=gray]{hyperref}

\usepackage{url}

\usepackage{enumerate}
\usepackage{todonotes} %
\usepackage{enumitem}  %

\usepackage{titlesec}

\usepackage{makecell}

\usepackage{pifont} %

\usepackage{algorithm,algpseudocode}

\usepackage{amsthm}
\usepackage{framed}
\definecolor{shadecolor}{gray}{0.95}
\newcolumntype{L}[1]{>{\raggedright\let\newline\\\arraybackslash\hspace{0pt}}m{#1}}
\newcolumntype{C}[1]{>{\centering\let\newline\\\arraybackslash\hspace{0pt}}m{#1}}
\newcolumntype{R}[1]{>{\raggedleft\let\newline\\\arraybackslash\hspace{0pt}}m{#1}}

\newcommand{\ignore}[1]{}

\makeatletter
\DeclareRobustCommand\onedot{\futurelet\@let@token\@onedot}
\def\@onedot{\ifx\@let@token.\else.\null\fi\xspace}

\makeatother

\definecolor{MyDarkBlue}{rgb}{0,0.08,1}
\definecolor{MyDarkGreen}{rgb}{0.02,0.6,0.02}
\definecolor{MyDarkRed}{rgb}{0.8,0.02,0.02}
\definecolor{MyDarkOrange}{rgb}{0.40,0.2,0.02}
\definecolor{MyPurple}{RGB}{111,0,255}
\definecolor{MyRed}{rgb}{1.0,0.0,0.0}
\definecolor{MyGold}{rgb}{0.75,0.6,0.12}
\definecolor{MyDarkgray}{rgb}{0.66, 0.66, 0.66}

\usepackage[utf8]{inputenc} %
\usepackage[T1]{fontenc}    %
\usepackage{natbib}

\title{Roadmap towards Meta-being}

\author{%
Tianyi Huang$^{1,2}$\\
\And
Stan Z. Li$^{1}$
\And
Xin Yuan$^{1,3}$
\And
Shenghui Cheng$^{1,2,3}$
\AND
$^1$ Westlake University, Hangzhou, China\\
$^2$ Westlake Institute for Advanced Study, Hangzhou, Chin\\
$^3$ Research Center for the Industries of the Future, Hangzhou, China
}

\begin{document}
\maketitle

\footnotetext{Correspondence to: Shenghui Cheng (chengshenghui@westlake.edu.cn)}

\vspace{-1em}
\begin{abstract}
\vspace{-0.5em}
Metaverse is a perpetual and persistent multi-user environment that merges physical reality with digital virtuality.
It is widely considered to be the next revolution of the Internet.
Digital humans are a critical part of Metaverse.
They are driven by artificial intelligence (AI) and deployed in many applications.
However, it is a complex process to construct digital humans which can be well combined with the Metaverse elements, such as immersion creation, connection construction, and economic operation.
In this paper, we present the roadmap of Meta-being to construct the digital human in Metaverse.
In this roadmap, we first need to model and render a digital human model for immersive display in Metaverse. 
Then we add a dialogue system with audio, facial expressions, and movements for this digital human.
Finally, we can apply our digital human in the fields of the economy in Metaverse with the consideration of security.
We also construct a digital human in Metaverse to implement our roadmap.
Numerous advanced technologies and devices, such as AI, Natural Language Processing (NLP), and motion capture, are used in our implementation.
This digital human can be applied to many applications, such as education and exhibition.
\end{abstract}
\vspace{-0.5em}
\section{Introduction}
\vspace{-0.5em}

Since the invention of the camera in 1839, photos have been used to present the appearances of great characters in history for nearly two hundred years.
However, photos cannot satisfy our desire to interact with someone.
With the development of computers, Metaverse which is widely considered to be the next revolution of the Internet makes this desire come to the truth~\cite{cheng2022roadmap}.

Metaverse is a perpetual and persistent multi-user environment with the mergence of physical reality with digital virtuality~\cite{loveys2022exploring,van2021metaverse,mystakidis2022metaverse}.
It is based on the convergence of technologies such as Virtual Reality (VR), Augmented Reality (AR), and Mixed Reality (MR)~\cite{azuma1997survey,costanza2009mixed,schuemie2001research}, and thus enables multi-sensory interactions with virtual environments, digital objects, and people.
Our desire to interact with someone can be implemented on the digital humans in Metaverse~\cite{fang2021metahuman,9757550}.
They are autonomously-animated virtual people driven by AI and deployed in many applications, such as healthcare, customer service, and education.
However, it is complex to construct a digital human which can be well combined with the elements in Metaverse, such as immersion creation, connection construction, and economic operation.

\begin{figure}[hbt]
    \centering
    \includegraphics[width=10cm]{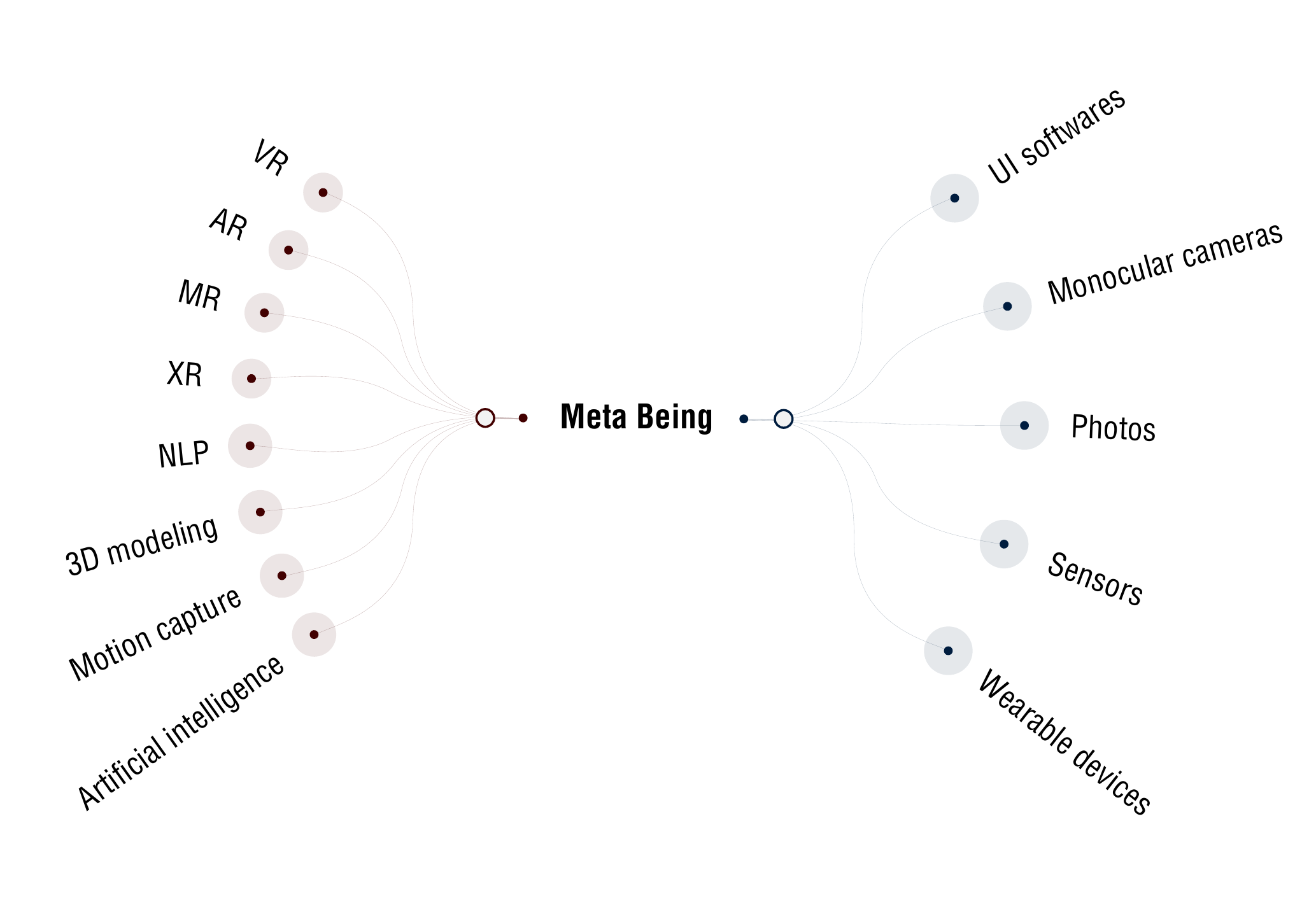}
    \caption{The technologies and devices in Meta-being.}
    \label{fig5}
\end{figure}

In this paper, we present the roadmap of Meta-being to construct the digital human in Metaverse.
In this roadmap, we first need to model and render a digital human model for immersive display in Metaverse. 
Then we add a dialogue system with audio, facial expressions, and movements for this digital human.
Finally, we can apply our digital human in the fields of the economy in Metaverse with the consideration of security.
In this way, we can appreciate the motions and speakings of digital humans and interact with them in Metaverse as in physical reality.
Meta-being is a complex process.
As we can see from Fig.~\ref{fig5}, many advanced technologies and devices are used to perform our Meta-being.

Our contributions are summarized as follows:
\begin{itemize}
    \item We present the roadmap for the Meta-being of digital humans.
    \item We construct a digital human to implement Meta-being.
\end{itemize}

The rest of this paper is organized as follows.
Section 2 reviews the Metaverse and digital human beings.
In Section 3, we give the roadmap of Meta-being.
In Section 4, we construct a digital human to implement Meta-being.
In the end, we conclude the paper in Section 5.

\vspace{-0.5em}
\section{Digital Human Beings in Metaverse}
\vspace{-0.5em}

The term Metaverse was invented and first appeared in Neal Stevenson’s science fiction novel Snow Crash published in 1992~\cite{stephenson2003snow}.
It represented a parallel virtual reality universe created from computer graphics to make users worldwide access and connect through goggles and earphones.
Digital humans as an important part of Metaverse are computer-generated and computer-controlled characters with a human-like appearance and can simulate human face-to-face interactions~\cite{magnenat2005virtual}. 
At the same time, with the development of artificial intelligence, intelligent digital human has become one of the new development directions.
Digital humans are often used in education, movies, games, and other fields. Each application domain requires different attributes at different levels, such as autonomous behaviour, natural language communication, recognition of real people, etc. Different intelligent decision-making techniques, such as artificial neural networks, need to be used to build a virtual human framework. Noma et al.~\cite{noma2000design} created a digital human presenter based on the JackTM animated agent system in 2000. The input to this system is in the form of spoken text with embedded commands (mostly related to the digital presenter's body language). 
The system then has the digital humans as a presenter with presentation skills in real-time 3D animation synchronized with the voice output. However, due to technical limitations, their digital human cannot synchronize lip movement with sound, and the movements are not smooth and natural enough. 
In 2012, Rizzo et al.~\cite{rizzo2011simcoach} developed a digital human, SimCoach, which could provide healthcare information and support, especially for military members, veterans, and their important ones.
Specifically, SimCoach breaks down barriers to online care by providing confidential assistance for military members and others in exploring and accessing healthcare content and facilitating on-site care. However, SimCoach's anthropomorphic degree is not high, and the interaction with users is insufficient. And he can only provide support for specific groups of people and specific medical problems. Due to technical limitations, these digital humans are not so much like virtual digital humans but like computer systems with human images. In recent years, a group of virtual people such as Miquela Sousa has appeared~\cite{Dalayli2020}. 
They exist in the network (e.g., social media), and the robot-focused representations are highly anthropomorphic and interactive.
Some researchers think although they live in the virtual world as humans live in the real world, with ``jobs", ``emotions", and ``thoughts",
they are neither real humans nor real phenomena, but over time digital humans will be considered human in social media, which is seen as a simulated environment~\cite{hubble2018miquela,marwick2018algorithmic}.
Though the ethical issues that come with it will be a considerable challenge to tackle.

\vspace{-0.5em}
\section{Meta-being}
\vspace{-0.5em}

\begin{figure}
    \centering
    \includegraphics[width=10cm]{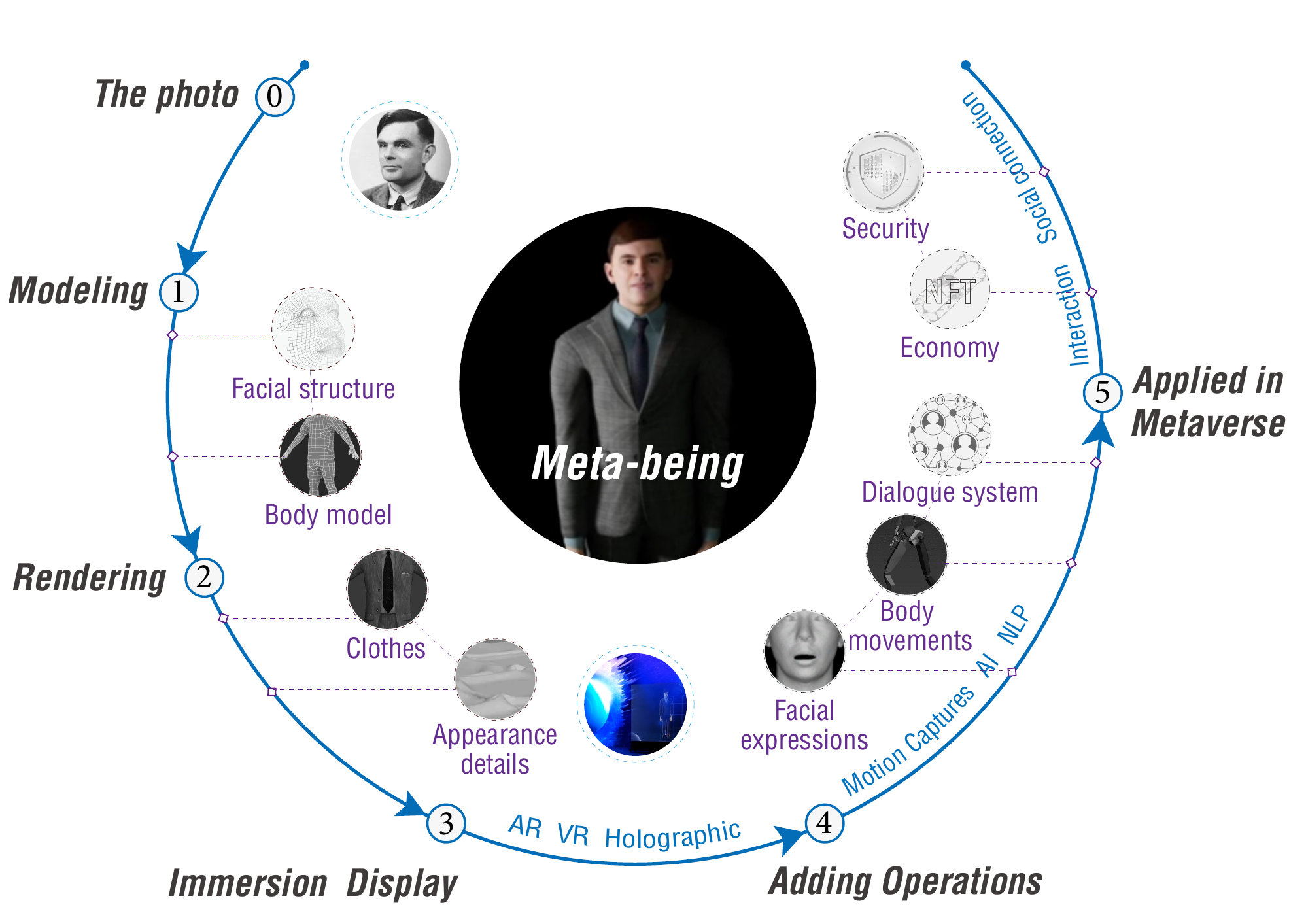}
    \caption{The roadmap towards Meta-being.}
    \label{fig8}
\end{figure}

The digital cyberspace of Metaverse includes immersion creation, hardware support, text interpretation, audio processing, connection construction, economy operation, and security protection~\cite{cheng2022roadmap}.
To create a digital human which can be well combined with the above elements, we propose the roadmap towards Meta-being as follows.
\begin{itemize}
    \item[1)] The body and the face of our digital human should be modeled.
    \item[2)] The resulting model should be rendered in detail.
    \item[3)] We should give an immersive display of our digital human in Metaverse by some display technologies, like VR and holographic projection. 
    \item[4)] We should generate a dialogue system with audio, facial expressions, and movements for this digital human.
    \item[5)] We can apply our digital human in the fields of the economy in Metaverse with the consideration of security.
\end{itemize}
As we can see, Meta-being is a complex process.
Many advanced technologies and devices, such as 3D modeling, AI, and VR, are used to perform our Meta-being.
This roadmap is demonstrated in Fig.~\ref{fig8}.

\subsection{Modeling and Rendering}
\begin{figure}[t]
\centering
  \includegraphics[width=10cm]{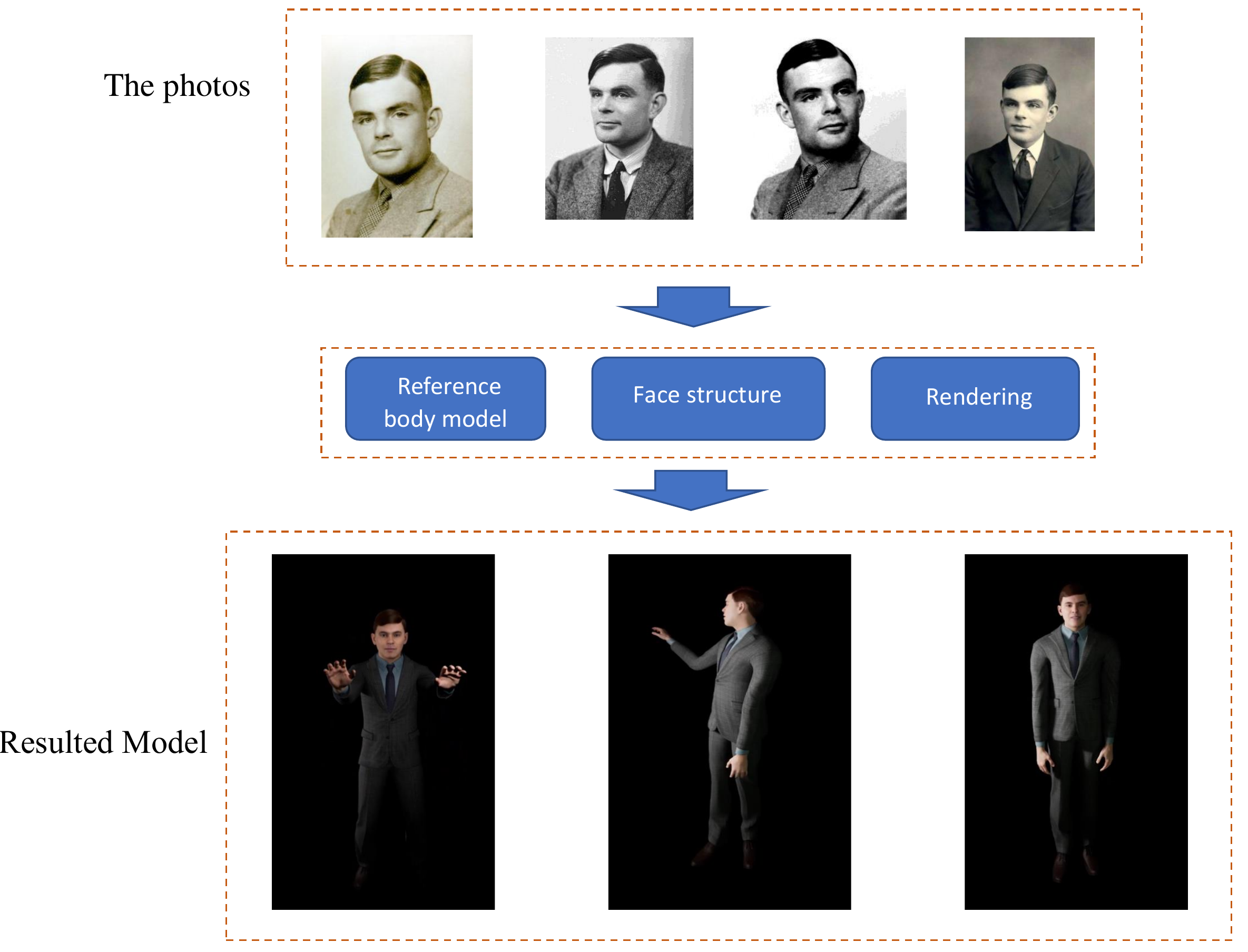}
  \caption{The modeling process and the resulting model.}
  \label{fig1}
\end{figure}

For a digital human in Meta-being, we should construct and then render a digital human model~\cite{demirel2007applications,zhang2022graphical,cheng2015data}.
To this end, we first need a human body model as a reference.
This model can be an adult or a child with an athletic body with relevant muscle mass and no significant body fat~\cite{azevedo2020towards}.
Some 3D scanning technologies, such as TEN24, can be used to capture the reference model~\cite{berdic2017creation,ratner20123}.
Second, we can use 3D Face Modeling technologies, like triangulation~\cite{morris2000image},3D Morphable Model~\cite{abrevaya2018multilinear}, to construct the face of our digital human by the reference photos in detail.
Finally, in the rendering step, we add the hair and clothes to our modeling digital human and fine-tune the appearance structures, such as the color and texture of the skin. 
In this modeling, the triangulation set, $T=\{t_1,t_2,...,t_n\}$ is along with a texture map set, $A=\{a_1,a_2,...,a_n\}$, on all of the face triangles.
The photo set $V'=\{v'_1,v'_2,...,v'_k\}$ taken with a calibrated camera in Metaverse can be computed as
\begin{eqnarray}
\label{eq1}
V' = f(T,A)
\end{eqnarray}
$a_i$ for a given triangulation $t_i$ can be estimated directly from the actual photo set, $V=\{v_1,v_2,...,v_k\}$, where $v_j$ is along with the same camera angle of $v'_j$, as follows. 
\begin{eqnarray}
A = f^{-1}(T,V)
\end{eqnarray}
Thus $V'$ from equation~\eqref{eq1} depends only on $T$ and $V$.
With the assumption that the photos along with the triangulation constraints are sufficient to define a unique face, our goal is to get the maximum Similarity $U$ for the actual photo set and the photo set from Metaverse as
\begin{eqnarray}
\arg\max_{t_i}\sum_{j=1}^{k}\sum_{i=1}^{n}U(v_j,v'_j|t_i)
\end{eqnarray}
Finally, in the rendering step, we add the hair and clothes to our modeling digital human of Turing and fine-tune the appearance structures, such as the color and texture of the skin. 
Fig.~\ref{fig1} shows the modeling process and the resulting model with different poses.

\subsection{Immersive Display}
For the immersion creation in Metaverse, immersive VR mainly presents virtual environments using cave automatic virtual environments or head-mounted displays, which can provide users with a strong and immersive feeling of presence in a 3D virtual world.
The immersive display of digital humans is a critical part of immersion creation in Metaverse for us to interact with digital humans.
Existing immersive display methods of digital humans are usually based on predefined elements, including body motions and facial expressions.
This could preclude the immersive feeling of human interaction with a digital human in the Metaverse since these elements may be different from the real ones with limited predefined elements.
To address this issue, we should capture many real scenes and use those images or videos to generate new ones and update the scenes in the metaverse within a short period of time, possibly in
real-time eventually.
Computational imaging may provide a promising solution to capture scenes efficiently in a low-cost, low-bandwidth manner with the help of deep learning~\cite{barbastathis2019use,mait2018computational}.

\subsection{Adding Operations}
For a Meta-being digital human, we also need to create the dialogue system, body movements, and facial expressions for our model.
The dialogue of our digital human is mainly based on our QA system generated by NLP algorithms~\cite{harabagiu2000falcon}.
This system first classifies the questions from users by automatic speech recognition and then extracts the relevant answers~\cite{yu2016automatic,sarkar2015nlp}.
The intent of the question could be understood by identifying the starting phrase or words of this question.
To answer the question, we can identify areas of interest from the voluminous content for this question and find the related paragraphs,
and then score each sentence in the related paragraphs by the number of proper nouns, the similarity between this sentence and the question, and so on.
A sentence can be considered a part of the answer if its score is above a particular threshold.

Human motion capture, like Xsens MVN motion capture, can be used to get the body motions and for digital human~\cite{schepers2018xsens}.
It tracks the motion of the human body defined by a biomechanical model consisting of some segments, like the neck, head, shoulders, feet, and so on.
For each body segment $B$, all kinematic quantities are expressed in a standard, local coordinate frame $L$, which is a right-handed Cartesian coordinate system.
Given the sensor $S$, we can obtain the position of each segment $^{L}p_{B}$ and corresponding orientation $^{LB}q$ by
\begin{eqnarray}
^{LB}q = ^{LS}q \otimes   ^{BS}q^{*}
\end{eqnarray}
and
\begin{eqnarray}
^{L}p_{B} = ^{L}p_{S}+ ^{LB}q \otimes ^{B}r_{BS} \otimes ^{LB}q^{*}
\end{eqnarray}
where $^{BS}q$ denotes the relative orientation of the sensor w.r.t. the body, $^{B}r_{BS}$ denotes the position of the sensor w.r.t. the segment origin expressed in the segment
frame, $\otimes$ denotes the quaternion multiplication, and $*$ denotes the complex conjugate of the quaternion.

\subsection{Applied in Metaverse}
The economic operation in Metaverse is for exchanging virtual goods through a blockchain digital system.
Digital humans are one type of the most important virtual goods in Metaverse.
The applications of Meta-being digital humans are driven by interaction and social connection.
For example, interaction with digital recreations of historical figures has existed for almost a decade, e.g., an interactive life-size video of Abraham Lincoln that is used at the National Civil War Museum.
In this case, education has become an important application of Meta-being digital humans.
By Meta-being digital humans with an artistic style, the children’s attention can be well attracted during the learning process. 
Also, proposing a Meta-being digital human curriculum with a ready-to-use software package would also improve/update the engineering quality and provide a unique educational experience for students~\cite{demirel2007applications}.

Widespread adoption of the Metaverse comes with many unique threats to user privacy and security, especially in the interaction with digital humans.
Biometric data reveals a host of personally identifiable information which can in turn be used to potentially manipulate users on a psychological level through the creation of avatars that are adaptable to user preferences~\cite{buck2022security}.
AI may have the capability of privacy protection through algorithms that automatically and dynamically detect user privacy references from diverse contexts in the Metaverse~\cite{zang2022evnet}.
An AI-powered security system will emerge in the Metaverse.

\section{The Implementation of Meta-being}
\begin{figure}
\centering
  \includegraphics[width=1.1in]{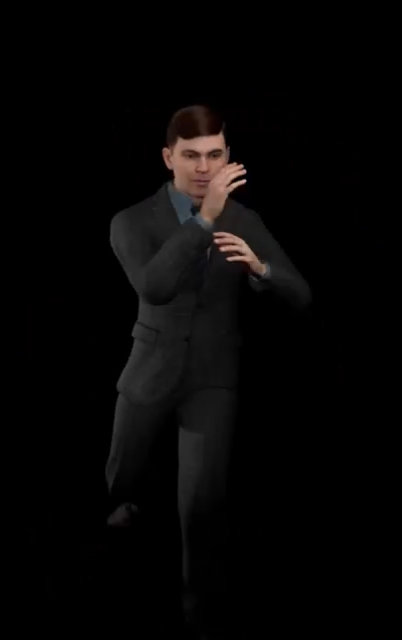}
   \includegraphics[width=1.1in]{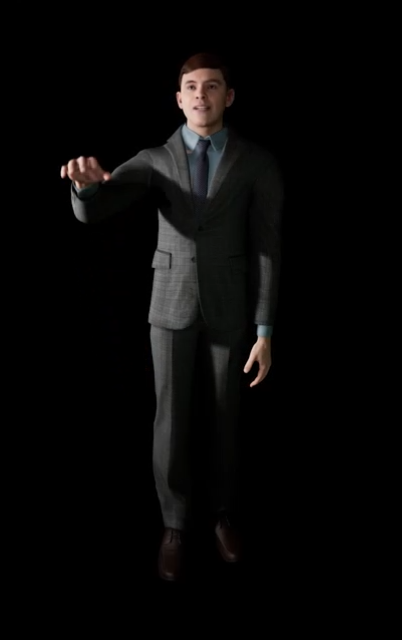}
   \includegraphics[width=1.1in]{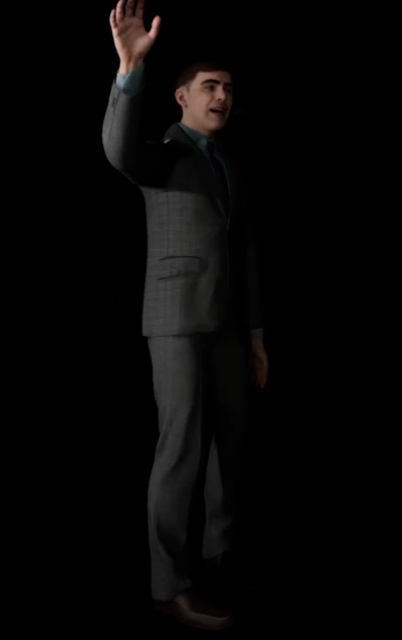}
  \includegraphics[width=1.1in]{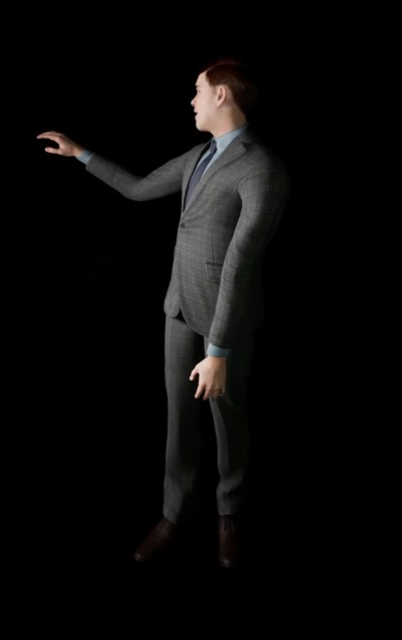}
  \caption{The actions of our digital Turing}
  \label{fig2}
\end{figure}

\begin{figure}
\centering
  \includegraphics[width=1.1in]{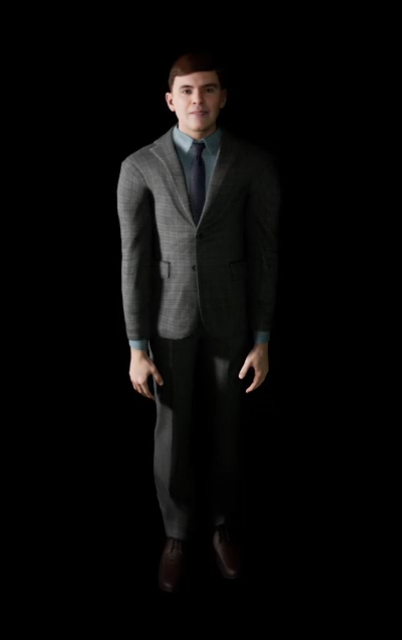}
  \includegraphics[width=1.1in]{s22.png}
  \includegraphics[width=1.1in]{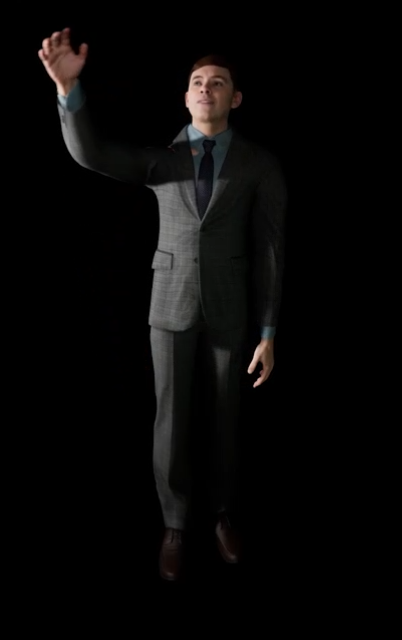}
  \includegraphics[width=1.1in]{s24.png}
  \caption{The action sequence of giving a class.}
  \label{fig3}
\end{figure}

\begin{figure}
\centering
  \includegraphics[width=1.1in]{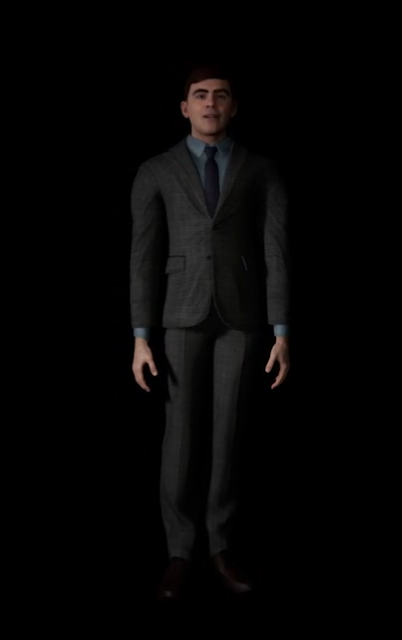}
  \includegraphics[width=1.1in]{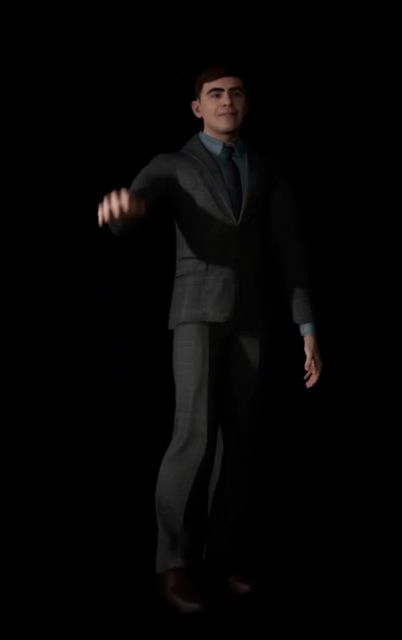}
  \includegraphics[width=1.1in]{s33.png}
  \includegraphics[width=1.1in]{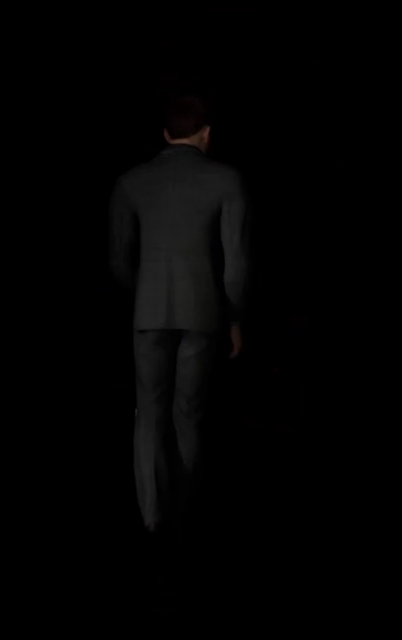}
  \caption{The action sequence of saying goodbye.}
  \label{fig4}
\end{figure}

As an implementation of Meta-being, we have made a digital human of Alan Mathison Turing by modeling the digital human model of Turing and then creating dialogue and movement functions, including body motions and facial expressions.
Then in Metaverse, we can appreciate the body motions, facial expressions, and speakings of digital Truing.
We also can talk to our digital Turing in Metaverse.
Our digital Truing can be applied to many applications, such as education and exhibition.
By mixing VR and AR, our digital Truing can explain computer knowledge to students and introduce new products to consumers in an immersive environment.
We display three interesting action sequences of our Turing model in Fig.~\ref{fig2}-\ref{fig4}.

\begin{figure}
    \centering
    \includegraphics[width=10cm]{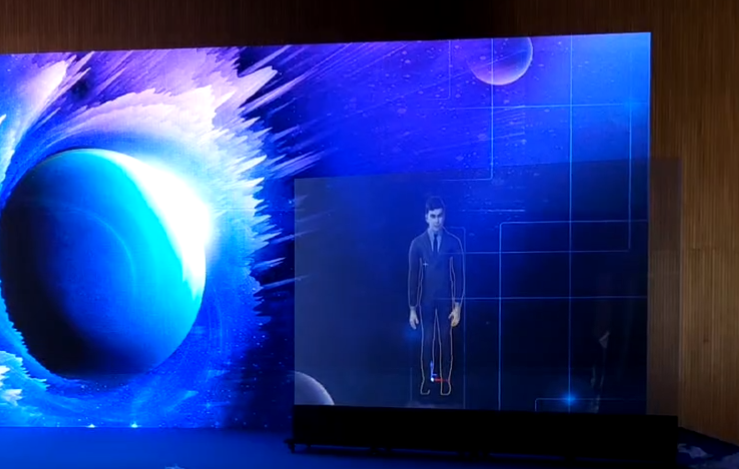}
    \caption{The holographic projection of our digital Turing.}
    \label{fig7}
\end{figure} 

Holographic projection technology breaks the traditional presentation of images by combining 3D technology and holographic technology~\cite{elmorshidy2010holographic,huebschman2003dynamic}.
It has become a hot application technology in the Metaverse for immersion display~\cite{aburbeian2022technology}.
By holographic projection, we have presented our digital Turing at some expositions.
Fig.~\ref{fig7} shows an example of our presentation.
As we can see, our vivid digital Turing is well integrated with the stage’s background.

\section{Conclusion}
In this paper, we have presented the roadmap for the Meta-beings to make digital humans can be better combined with the elements in Metaverse.
In this roadmap, we model the body and face of a digital human in a 3D space.
Then, in the rendering step, we add clothes and appearance details to our model.
We also need to create a dialogue system, body movements, and facial expressions for this model.
In this manner, we can appreciate its motions and speaking and interact with him in Metaverse as in physical reality.
It is a complex process to implement our Meta-being on a digital human.
Many advanced technologies and devices, such as AI, NLP, and motion capture, are used to perform our Meta-being.
As an example, we implement our Meta-being on Turing, a great scientist of computer science.

{

\bibliographystyle{plainnat}
\bibliography{reference}
}

\end{document}